\documentclass[altaffilletter,aps,nofootinbib,twocolumn,prd,eqsecnum,preprintnumbers,superscriptaddress]{revtex4-1}
\pdfoutput=1
\usepackage[caption=false]{subfig}
\usepackage{graphicx}
\usepackage{amsmath}
\usepackage{amsfonts}
\usepackage{amssymb}
\usepackage{color}
\usepackage{bm}
\usepackage{mathrsfs}
\usepackage{epstopdf}
\usepackage{url}
\usepackage{footnote}
\usepackage{textcomp}
\usepackage{dsfont}
\usepackage{ulem}
\usepackage{hyperref}
\usepackage{enumerate}

\makeatletter
\newcommand*{\rom}[1]{\expandafter\@slowromancap\romannumeral #1@}
\makeatother

\begin{document}

\preprint{CERN-TH-2021-011}

\title{Black hole quasinormal modes and isospectrality in Deser-Woodard nonlocal gravity}

\author{Che-Yu Chen}
\email{b97202056@gmail.com} 
\affiliation{Institute of Physics, Academia Sinica, Taipei, Taiwan 11529}

\author{Sohyun Park}
\email{sohyun.park@cern.ch}
\affiliation{Theoretical Physics Department, CERN, CH-1211 Gen\`eve 23, Switzerland}

\begin{abstract}
We investigate the gravitational perturbations of the Schwarzschild black hole in the nonlocal gravity model recently proposed by Deser and Woodard (DW-II). The analysis is performed in the localized version in which the nonlocal corrections are represented by some auxiliary scalar fields. We find that the nonlocal corrections do not affect the axial gravitational perturbations, and hence the axial modes are completely identical to those in General Relativity (GR). However, the polar modes get different from their GR counterparts when the scalar fields are excited at the background level. In such a case the polar modes are sourced by an additional massless scalar mode and, as a result, the isospectrality between the axial and the polar modes breaks down. We also perform a similar analysis for the predecessor of this model (DW-I) and arrive at the same conclusion for it.
\end{abstract}

\maketitle

\section{Introduction}   

The direct detection of gravitational waves emitted from binary black hole mergers \cite{Abbott:2016blz,LIGOScientific:2018mvr,Abbott:2020niy} ushers in an exciting era of gravitational wave astronomy. More surprisingly, the current observing sensitivity already allows the detection of roughly one candidate event per week in average \cite{Abbott:2020niy}, not to mention the amount of
data available with the further upgrades and improvements of the detectors in the future. The gravitational waves emitted from these events carry useful information about the spacetime under strong gravitational fields. Therefore, such a detection facilitates us to probe the physics in the extreme environments such as black holes, and to test whether our current understanding of black holes and gravity based on general relativity (GR) is correct or not.   

Essentially, the gravitational wave signals associated with a binary black hole merger consist of three stages. The first is the inspiral stage during which the binary black holes rotate around each other. At this stage, the gravitational waves can be modeled well by the post-Newtonian approaches. The second is the merger stage, in which the gravitational fields are so strong that purely numerical analysis is inevitable. The third stage, which is called the ringdown, corresponds to the post-merger phase of the event. At this stage, the binary black holes have coalesced, and the distortion of the final black hole is gradually settling. The black hole during this ringdown stage loses its energy by emitting gravitational waves, with frequencies governed by a superposition of decaying oscillations, i.e., the quasinormal modes (QNMs) \cite{Kokkotas:1999bd,Berti:2009kk,Konoplya:2011qq}. The ringing black hole and the associated QNMs can be described well by black hole perturbation theories. An important realization is that the QNM spectrum only depends on the parameters quantifying the black hole, such as mass, spin, and whatever parameters that characterize the black hole in a particular gravitational theory. For instance, the QNM spectrum of a Kerr black hole in GR is completely determined by its mass and spin. The spectrum does not depend on what initially drives the ringings. This means that by examining the QNM frequencies one can test the no-hair theorem in the underlying theory of gravity. That is, the QNM spectrum is an observable to distinguish between GR and other gravitational theories. In this respect, the investigation of black hole perturbations and the associated QNM spectra in the context of theories beyond GR has been an intensive field of research, see for instance Refs.~\cite{Kobayashi:2012kh,Kobayashi:2014wsa,Blazquez-Salcedo:2016enn,Bhattacharyya:2017tyc,Glampedakis:2017dvb,Chen:2018mkf,Chen:2018vuw,Chen:2019iuo,Moulin:2019ekf,Chen:2020evr}.

In addition to the no-hair theorem, there are other fundamental properties which, by examining black hole QNMs, can be used to discriminate GR from other theories. For example, it is well-known that in the eikonal approximation, the QNM frequencies for stationary, axisymmetric, and asymptotically flat black holes in GR are tightly related to the properties of the unstable photon orbits around the black hole \cite{Cardoso:2008bp,Dolan:2010wr,Yang:2012he}. The eikonal QNMs are also related to the apparent size of black hole shadows \cite{Cuadros-Melgar:2020kqn}. Therefore, apart from extracting the mode frequencies and comparing the spectra with those in GR, one can test gravitational theories by examining these properties. Any breaking of them would be a smoking-gun for different physics beyond Einstein's gravity \cite{Konoplya:2017wot,Churilova:2019jqx,Glampedakis:2019dqh,Chen:2019dip}. 

Another important feature of QNMs in GR is the isospectrality. For the Schwarzschild black hole in GR, the gravitational perturbations of the axial (parity-odd) \cite{Regge:1957td} and the polar (parity-even) modes \cite{Zerilli:1970se} share an identical spectrum, even though their master equations are completely different from each other \cite{Chandrabook}. This interesting property also holds for the Reissner-Nordstr\"om, Kerr, as well as Kerr-Newman (to linear-order in the spin) black holes \cite{Pani:2013ija,Pani:2013wsa}. It was discovered also by Chandrasekhar \cite{Chandrabook}, in particular for Schwarzschild black holes, that the master equations governing these two modes are related by a specific transformation. It was not realized until very recently \cite{Glampedakis:2017rar} that the transformation introduced in Ref.~\cite{Chandrabook} is actually a particular subclass of the Darboux transformation \cite{Darboux:1882}.  The identification of this relation and its generalization \cite{Glampedakis:2017rar,Yurov:2018ynn} helps to investigate the isospectrality of more complex systems, such as the Kerr black hole \cite{Teukolsky:1973ha,Sasaki:1981sx}. In fact, the isospectrality of black hole QNMs is a longstanding issue \cite{Moulin:2019bfh}. For example, the physical origin of this property is unclear. In addition, although the isospectrality holds for most of the astrophysically relevant black holes in GR, not all black hole solutions in GR share this property \cite{Ferrari:2000ep,Cardoso:2001bb,Berti:2003ud,Blazquez-Salcedo:2019nwd} and the range of its viability is still an open question. In many theories beyond GR, such as $f(R)$ gravity \cite{Bhattacharyya:2017tyc,Bhattacharyya:2018qbe,Datta:2019npq}, dynamical Chern-Simons gravity \cite{Bhattacharyya:2018hsj}, and scalar-tensor theories \cite{Kobayashi:2012kh,Kobayashi:2014wsa}, the isospectrality is generically broken. Furthermore, it has been shown recently using a parameterization approach that the isospectrality is actually very fragile \cite{Cardoso:2019mqo}. Therefore, any observational indication of the isospectrality breaking would strongly hint toward new physics beyond GR \cite{Shankaranarayanan:2019yjx}.

In this work, we will focus on the possibility of testing gravitational theories by examining the isospectrality of black hole QNMs. In particular, we will investigate the axial and the polar gravitational perturbations of the Schwarzschild black hole in the Deser-Woodard nonlocal gravity models I and II (DW-I  \cite{Deser:2007jk} and DW-II \cite{Deser:2019lmm}). This class of nonlocal gravity models was proposed by Deser and Woodard \cite{Deser:2007jk, Deser:2019lmm} with the motivation that quantum loop effects of infrared gravitons abundant in the early universe are essentially nonlocal and the effects might persist in the late universe \cite{Woodard:2014iga, Woodard:2018gfj}. Deser and Woodard particularly aimed at a phenomenological model that can generate the current phase of cosmic acceleration without invoking dark energy.\footnote{Other classes of nonlocal gravity models with the same goal of reproducing the late-time acceleration without dark energy have been proposed and studied extensively \cite{Barvinsky:2011hd, Maggiore:2013mea, Maggiore:2014sia, Vardanyan:2017kal, Amendola:2017qge, Tian:2018bmn}.} 
Their first model DW-I  \cite{Deser:2007jk} had been very successful in various aspects. It reproduced the background expansion identical to 
that of the $\Lambda$CDM expansion (without $\Lambda$) in GR \cite{Deffayet:2009ca}. In addition, once the background was fixed, the large scale structure formation predicted by the model was comparable to the one in GR \cite{Park:2012cp, Dodelson:2013sma, Park:2016jym, Nersisyan:2017mgj, Park:2017zls, Amendola:2019fhc}. Even though the localized version \cite{Nojiri:2007uq} of this model possesses a scalar ghost \cite{Zhang:2016ykx, Nojiri:2010pw, Zhang:2011uv}, the constraint of the model precludes explosive growth of the ghost \cite{Park:2019btx}. 
However a recent analysis showed that DW-I violates the solar system constraints \cite{Belgacem:2018wtb}. Deser and Woodard then proposed an improved model (DW-II) \cite{Deser:2019lmm} that might pass the solar system constraints. Therefore, in this paper we focus on the DW-II gravity and briefly remark for the DW-I gravity at the end.

For the Schwarzschild QNMs in the DW-II model, we first find that the axial modes are not affected by the nonlocal corrections. Furthermore, due to the special role played by the nonlocal term that whether the auxiliary scalar fields are excited or not depends on the boundary conditions, we find that the master equation of the polar modes is sourced by an additional scalar mode corresponding to the auxiliary scalar fields. However, this scalar mode is not excited as long as the auxiliary scalar fields are quiescent at the unperturbed level. If the scalar mode is not excited, the QNMs of the axial and the polar modes reduce to those in GR and the isospectrality is preserved, otherwise the isospectrality gets broken. As a remark, due to the fact that the DW-I and DW-II models are similar in their mathematical structures, the same conclusion can be drawn for the DW-I model.

The rest of this paper is organized as follows. In Section~\ref{sec.2}, we briefly review the DW-II nonlocal gravity
including its action, equations of motion, and localization. In Section~\ref{sec.3}, we discuss the Schwarzschild black hole in the DW-II nonlocal gravity. Then, we perturb the Schwarzschild black hole in Section~\ref{sec.IV}, and derive the master equations of the axial and polar perturbations. The results will be compared with those in GR. In Section~\ref{sec.dwI}, we perform a similar analysis in the framework of DW-I nonlocal gravity. Finally, Section~\ref{sec.conclus} draws our conclusions and discussions.

\section{Deser-Woodard-II nonlocal gravity}\label{sec.2}

The action of the DW-II nonlocal gravity is \cite{Deser:2019lmm}
\begin{equation}
\mathcal{S}_{\textrm{DW-II}}=\frac{1}{16\pi}\int d^4x\sqrt{-g}R\left[1+f\left(\zeta\right)\right]+\mathcal{S}_m\,,\label{actionDW2}
\end{equation}
where $\mathcal{S}_m$ is the matter action. The nonlocal distortion function $f$ is an arbitrary function of the inverse scalar d'Alembertian acting on $(\partial\phi)^2\equiv g^{\mu\nu}\partial_\mu\phi\partial_\nu\phi$, where $\phi$ is the inverse scalar d'Alembertian acting on the Ricci scalar $R$. More precisely, the scalar fields $\zeta$ and $\phi$ can be written explicitly as follows:
\begin{align}
\zeta&\equiv\Box^{-1}\left(\partial\phi\right)^2\,,\label{defzeta}\\
\phi&\equiv\Box^{-1}R\,.\label{defphi}
\end{align}
The distortion function $f$ is thus a function of $\zeta$. The presence of the distortion function adds nonlocal corrections to the Einstein-Hilbert action. Note that we have assumed $c=G=1$ in this paper, where $c$ is the speed of light and $G$ is the gravitational constant.

As has been shown in Ref.~\cite{Deser:2019lmm}, the DW-II model can be localized by introducing two additional Lagrange multipliers $\Xi$ and $\psi$ as follows{\footnote{The notations for the auxiliary scalar fields used in Ref.~\cite{Deser:2019lmm} ($X,Y,U,V$) and ours ($\phi,\zeta,\Xi,\psi$) are different. They can be directly converted as follows: $X\rightarrow \phi$, $Y\rightarrow\zeta$, $U\rightarrow-\Xi$, and $V\rightarrow-\psi$.}}
\begin{align}
\mathcal{S}_{\textrm{DW-II}}&=\frac{1}{16\pi}\int d^4x\sqrt{-g}\Big[R\left(1+f-\Xi\right)-\partial_\mu\Xi\partial^\mu\phi\nonumber\\&-\psi\left(\partial\phi\right)^2-\partial_\mu\psi\partial^\mu\zeta\Big]+\mathcal{S}_m\,.\label{actionDW2}
\end{align}
The equations of motion are derived by varying the action \eqref{actionDW2} with respect to $\Xi$, $\psi$, $\phi$, $\zeta$, and the metric $g_{\mu\nu}$. Varying the action with respect to the Lagrange multipliers $\Xi$ and $\psi$ gives
\begin{align}
\Box\phi&=R\,,\label{phiboxR}\\
\Box\zeta&=\left(\partial\phi\right)^2\,,\label{zetaphisquare}
\end{align}
which correspond to the definition of the scalar fields $\phi$ and $\zeta$ given by Eqs.~\eqref{defphi} and \eqref{defzeta}, respectively. Then, the variation with respect to the scalar fields $\phi$ and $\zeta$ gives
\begin{align}
\Box\Xi&=-2\nabla_\mu\left(\psi\nabla^\mu\phi\right)\,,\label{Xi2psiphi}\\
\Box\psi&=-R\frac{df}{d\zeta}\,.\label{psiRdz}
\end{align}
Finally, the equation of motion derived by varying the action with respect to the metric reads
\begin{align}
&\left(G_{\mu\nu}+g_{\mu\nu}\Box-\nabla_\mu\nabla_\nu\right)\left(1+f-\Xi\right)\nonumber\\
&+\frac{1}{2}g_{\mu\nu}\left[\partial_\alpha\Xi\partial^\alpha\phi+\partial_\alpha\psi\partial^\alpha\zeta+\psi\left(\partial\phi\right)^2\right]\nonumber\\
&-\partial_{(\mu}\Xi\partial_{\nu)}\phi-\partial_{(\mu}\psi\partial_{\nu)}\zeta-\psi\partial_\mu\phi\partial_\nu\phi=8\pi T_{\mu\nu}\,,\label{graveqDW2}
\end{align}
where $G_{\mu\nu}$ and $T_{\mu\nu}$ stand for the Einstein tensor and the energy-momentum tensor, respectively.

\section{Schwarzschild spacetime in DW-II gravity}\label{sec.3}

Before investigating black hole perturbations in the DW-II nonlocal gravity, it is necessary to specify the background spacetime that is going to be perturbed. In this paper, we are going to consider the perturbations of the Schwarzschild black hole in the DW-II nonlocal gravity. In this section, we will use the requirement that the Schwarzschild metric is an exact vacuum solution ($T_{\mu\nu}=0$) of the DW-II nonlocal gravity to constrain the expressions of the auxiliary scalar fields. By doing so, we thus exhibit that the Schwarzschild solution does satisfy the field equations of the DW-II nonlocal gravity.

We consider the Schwarzschild metric
\begin{equation}
ds^2=-e^{2\bar\nu}dt^2+e^{-2\bar\nu}dr^2+r^2d\Omega_2^2\,,
\end{equation}
where $e^{2\bar\nu}=1-2m/r$ with $m$ being the mass of the black hole. From now on, quantities with a bar represent the quantities at the background level, in order to distinguish them from their linear perturbations. Since the Ricci scalar $R$ is identically zero, the scalar field equations \eqref{phiboxR} and \eqref{psiRdz} can be written as
\begin{align}
\bar\phi_{,rr}+2\left(\frac{1}{r}+\bar\nu_{,r}\right)\bar\phi_{,r}&=0\,,\label{sssphis}\\
\bar\psi_{,rr}+2\left(\frac{1}{r}+\bar\nu_{,r}\right)\bar\psi_{,r}&=0\,.\label{ssspsis}
\end{align}
Note that we have assumed that the scalar fields are functions of $r$ only at the background level. The above equations \eqref{sssphis} and \eqref{ssspsis} can be solved to get
\begin{equation}
\bar\phi_{,r}=C_\phi\frac{e^{-2\bar\nu}}{r^2}\,,\qquad \bar\psi_{,r}=C_\psi\frac{e^{-2\bar\nu}}{r^2}\,,\label{phipsiBGDW2}
\end{equation}
where $C_\phi$ and $C_\psi$ are integration constants. Furthermore, using Eqs.~\eqref{zetaphisquare} and \eqref{Xi2psiphi}, the other two scalar fields can be solved as
\begin{align}
\bar\zeta_{,r}&=\left(C_\phi\bar\phi+C_\zeta\right)\frac{e^{-2\bar\nu}}{r^2}\,,\label{zetabgdw2dd}\\
\bar\Xi_{,r}&=\left(-2C_\phi\bar\psi+C_\Xi\right)\frac{e^{-2\bar\nu}}{r^2}\,,\label{Xibgdw2dd}
\end{align}
where $C_\zeta$ and $C_\Xi$ are integration constants as well.

To proceed, we define $K\equiv f-\Xi$ and $\bar{K}\equiv \bar{f}-\bar\Xi$. Using the gravitational equation \eqref{graveqDW2}, we find that
\begin{align}
&\bar{K}=\textrm{constant}\,,\label{constraintK}\\
-C_\phi^2\bar\psi+C_\phi C_\psi\bar\phi&+C_\phi C_\Xi+C_\psi C_\zeta=0\,,\label{constraintDW2}
\end{align}
in which the later can be treated as a constraint to be satisfied by the integration constants. Since $\bar\phi$ and $\bar\psi$ satisfy Eq.~\eqref{phipsiBGDW2}, the constraint Eq.~\eqref{constraintDW2} can be rewritten as
\begin{equation}
C_\phi C_\Xi+C_\psi C_\zeta=C_{\infty}\,,
\end{equation}
where $C_{\infty}$ is a constant, whose value is determined by the asymptotic value of $\bar\phi$ and $\bar\psi$.{\footnote{In fact, the asymptotic values of the scalar fields may be set to zero in order to satisfy the asymptotic flatness condition. Note that at the weak-field regime of the DW-I model, the values of the scalar fields, as well as that of the distortion function $\bar f$, are supposed to approach zero asymptotically \cite{Chu:2018mld}. We expect this conclusion is also true for the DW-II model.}} 

Essentially, if any of the scalar fields ($\phi$, $\psi$, $\Xi$, $\zeta$) is excited at the background level, it could be a varying function of $r$. Whether the field is excited or not is determined by the value of the associated integration constants ($C_\phi$, $C_\psi$, $C_\Xi$, $C_\zeta$). We will show explicitly later how the excited fields are related to the black hole perturbations within the DW-II nonlocal gravity.

\section{Black hole perturbations}\label{sec.IV}
In this section, we will investigate the gravitational perturbations of the Schwarzschild black hole in the DW-II nonlocal gravity. Without loss of generality, the perturbed spacetime can be described by a non-stationary and axisymmetric metric in which the symmetrical axis is turned in such a way that no $\varphi$ (the azimuthal angle) dependence appears in the metric functions. In general, the metric can be written as follows \cite{Chandrabook}:
\begin{align}
ds^2=&-e^{2\nu}\left(dx^0\right)^2+e^{2\mu_1}\left(dx^1-\sigma dx^0-q_2dx^2-q_3dx^3\right)^2\nonumber\\&+e^{2\mu_2}\left(dx^2\right)^2+e^{2\mu_3}\left(dx^3\right)^2\,.\label{metricg}
\end{align}
Up to their first-order perturbations, the metric functions
\begin{align}
e^{2\nu}&=e^{2\bar\nu}\left(1+2\delta\nu\right)\,,\nonumber\\
e^{2\mu_1}&=r^2\sin^2\theta\left(1+2\delta\mu_1\right)\,,\nonumber\\
e^{2\mu_2}&=e^{-2\bar\nu}\left(1+2\delta\mu_2\right)\,,\nonumber\\
e^{2\mu_3}&=r^2\left(1+2\delta\mu_3\right)\,,\label{fielddefin}
\end{align}
are functions of time $t$ ($t=x^0$), radial coordinate $r$ ($r=x^2$), and polar angle $\theta$ ($\theta=x^3$). As we have mentioned, the perturbed metric is axisymmetric, therefore, we will assume that the metric functions are independent of the azimuthal angle $\varphi$ ($\varphi=x^1$). The quantities with a delta denote the corresponding first-order terms of the fields, and contribute to the polar perturbations in general. On the other hand, the functions $\sigma$, $q_2$, and $q_3$ are also functions of $t$, $r$, and $\theta$, and they correspond to the axial perturbations of the metric. We will discuss these two types (the polar and axial types) of perturbations in more detail later. In addition, the fields that induce nonlocal corrections should be perturbed, hence we have $K=\bar K+\delta K$, $\phi=\bar\phi+\delta\phi$, $\psi=\bar\psi+\delta\psi$, $\Xi=\bar\Xi+\delta\Xi$, and $\zeta=\bar\zeta+\delta\zeta$. We would like to emphasize again that the background spacetime that we are considering is the Schwarzschild metric where $e^{2\bar\nu}=1-2m/r$, and the scalar fields at the background level are functions of $r$ only. The field $\bar K$ is a constant, as shown in Eq.~\eqref{constraintK}.

\subsection{Tetrad formalism}

To study the perturbations of the spacetime metric \eqref{metricg}, we use the tetrad formalism in which one defines a basis associated with the metric \eqref{metricg} \cite{Chandrabook}:
\begin{align}
e^{\mu}_{(0)}&=\left(e^{-\nu},\quad\sigma e^{-\nu},\quad0,\quad0\right)\,,\nonumber\\
e^{\mu}_{(1)}&=\left(0,\quad e^{-\mu_1},\quad 0,\quad0\right)\,,\nonumber\\
e^{\mu}_{(2)}&=\left(0,\quad q_2e^{-\mu_2},\quad e^{-\mu_2},\quad0\right)\,,\nonumber\\
e^{\mu}_{(3)}&=\left(0,\quad q_3e^{-\mu_3},\quad 0,\quad e^{-\mu_3}\right)\,,\label{tetradbasis111}
\end{align}
and
\begin{align}
e_{\mu}^{(0)}&=\left(e^{\nu},\quad0,\quad0,\quad0\right)\,,\nonumber\\
e_{\mu}^{(1)}&=\left(-\sigma e^{\mu_1},\quad e^{\mu_1},\quad -q_2e^{\mu_1},\quad -q_3e^{\mu_1}\right)\,,\nonumber\\
e_{\mu}^{(2)}&=\left(0,\quad 0,\quad e^{\mu_2},\quad0\right)\,,\nonumber\\
e_{\mu}^{(3)}&=\left(0,\quad 0,\quad 0,\quad e^{\mu_3}\right)\,,\label{tetradbasis222}
\end{align}
where the tetrad indices are enclosed in parentheses to distinguish them from the tensor indices. The tetrad basis then satisfies
\begin{align}
e_{\mu}^{(a)}e^{\mu}_{(b)}&=\delta^{(a)}_{(b)}\,,\quad e_{\mu}^{(a)}e^{\nu}_{(a)}=\delta^{\nu}_{\mu}\,,\nonumber\\
e_{\mu}^{(a)}&=g_{\mu\nu}\eta^{(a)(b)}e^{\nu}_{(b)}\,,\nonumber\\
g_{\mu\nu}&=\eta_{(a)(b)}e_{\mu}^{(a)}e_{\nu}^{(b)}\equiv e_{(a)\mu}e_{\nu}^{(a)}\,.
\end{align}
Essentially, in the tetrad formalism we project the relevant quantities defined on the coordinate basis of $g_{\mu\nu}$ onto a chosen basis of $\eta_{(a)(b)}$ by constructing the tetrad basis correspondingly. Usually, $\eta_{(a)(b)}$ is assumed to be the Minkowskian matrix
\begin{equation}
\eta_{(a)(b)}=\eta^{(a)(b)}=\textrm{diag}\left(-1,1,1,1\right)\,.
\end{equation}
In this regard, any vector or tensor field can be projected onto the tetrad frame, in which the field can be expressed through its tetrad components:
\begin{align}
A_{\mu}&=e_{\mu}^{(a)}A_{(a)}\,,\quad A_{(a)}=e_{(a)}^{\mu}A_{\mu}\,,\nonumber\\
B_{\mu\nu}&=e_{\mu}^{(a)}e_{\nu}^{(b)}B_{(a)(b)}\,,\quad B_{(a)(b)}=e_{(a)}^{\mu}e_{(b)}^{\nu}B_{\mu\nu}\,.
\end{align}
One should notice that in the tetrad formalism, the covariant (partial) derivative in the original coordinate frame is replaced by the intrinsic (directional) derivative in the tetrad frame. For instance, the derivatives of an arbitrary rank-two object $H_{\mu\nu}$ in the two frames are related as follows \cite{Chandrabook}
\begin{align}
&\,H_{(a)(b)|(c)}\equiv e^{\lambda}_{(c)}H_{\mu\nu;\lambda}e_{(a)}^{\mu}e_{(b)}^{\nu}\nonumber\\
=&\,H_{(a)(b),(c)}\nonumber\\&-\eta^{(m)(n)}\left(\gamma_{(n)(a)(c)}H_{(m)(b)}+\gamma_{(n)(b)(c)}H_{(a)(m)}\right)\,,\label{2.7}
\end{align}
where a vertical rule and a comma denote the intrinsic and directional derivative with respect to the tetrad indices, respectively. A semicolon denotes a covariant derivative with respect to the tensor indices. Furthermore, the Ricci rotation coefficients are defined by
\begin{equation}
\gamma_{(c)(a)(b)}\equiv e_{(b)}^{\mu}e_{(a)\nu;\mu}e_{(c)}^{\nu}\,,
\end{equation}
and their components corresponding to the metric \eqref{metricg} are given in Ref.~\cite{Chandrabook}.

In the tetrad frame, the gravitational equation \eqref{graveqDW2} can be written as
\begin{widetext}
\begin{align}
R_{(a)(b)}\left(1+K\right)-e_{(a)}^\mu\left(K_{,(b)}\right)_{,\mu}+\gamma_{(c)(b)(a)}K_{,(d)}\eta^{(c)(d)}&+\eta_{(a)(b)}\left\{\Box K-\frac{R}{2}\left(1+K\right)+\frac{1}{2}\left[\partial_\rho\Xi\partial^\rho\phi+\partial_\rho\psi\partial^\rho\zeta+\psi\left(\partial\phi\right)^2\right]\right\}\nonumber\\
&-e^\rho_{((a)}e^\lambda_{(b))}\left(\partial_\rho\Xi\partial_\lambda\phi+\partial_\rho\psi\partial_\lambda\zeta+\psi\partial_\rho\phi\partial_\lambda\phi\right)=0\,.\label{greqtetrad}
\end{align}
\end{widetext}
Note that we have assumed a vacuum spacetime: $T_{\mu\nu}=0$.

\subsection{Axial perturbations}
According to how they react with respect to the parity change, the gravitational perturbations of a spherically symmetric black hole can be classified into two parts: the axial perturbations and the polar perturbations. The master equation governing the axial perturbations can be derived by perturbing Eq.~\eqref{greqtetrad} and considering its ($1,3$) and $(1,2)$ components (or equivalently the ($\varphi,\theta$) and $(\varphi,r)$ components).

Since the perturbed metric is axisymmetric and $\bar K$ is a constant, one can easily verify that the master equation of the axial perturbations $\Psi_{\textrm{aixal}}$ is described by 
\begin{equation}
R_{(1)(2)}=R_{(1)(3)}=0\,.
\end{equation}
Therefore, the nonlocal terms do not contribute to the axial perturbations, and the master equation is the same as the one for the Schwarzschild black hole in GR:
\begin{equation}
\left(\frac{d^2}{dr_*^2}+\omega^2\right)\Psi_{\textrm{aixal}}=V_{RW}\Psi_{\textrm{aixal}}\,,\label{axialmaster}
\end{equation}
where $r_*$ is the tortoise radius and $\omega$ is the quasinormal frequencies. The effective potential $V_{RW}$, or the so-called Regge-Wheeler potential, is defined as
\begin{equation}
{V_{RW}=\frac{e^{2\bar\nu}}{r^2}\left[l(l+1)-\frac{6m}{r}\right]}\,,\quad e^{2\bar\nu}=1-\frac{2m}{r}\,,
\end{equation}
where $l$ is the multipole number. The master equation \eqref{axialmaster} is equivalent to the Regge-Wheeler equation in GR \cite{Regge:1957td}, which describes the axial perturbations of the Schwarzschild black hole. Therefore, we conclude that the nonlocal terms do not change the axial modes.

\subsection{Polar perturbations}

In GR, the polar mode $\Psi_{\textrm{polar}}$ of the spacetime metric \eqref{metricg} is essentially a linear combination of the metric perturbations $\delta\nu$, $\delta\mu_1$, $\delta\mu_2$, and $\delta\mu_3$. In the DW-II nonlocal gravity, the polar mode may also depend on the perturbations of the nonlocal terms, such as $\delta K$, $\delta\phi$, $\delta\psi$, $\delta \Xi$, and $\delta\zeta$. In this subsection, we will use the tetrad formalism to derive the master equation of the polar perturbations for the Schwarzschild black hole in the DW-II nonlocal gravity. After obtaining the master equation, one can see how the nonlocal terms affect the polar modes. Regarding the derivation of the master equations of the Schwarzschild black hole in GR using the tetrad formalism, we refer the readers to Ref.~\cite{Chandrabook}.

First, the $(0,2)$ component of the linearized Eq.~\eqref{greqtetrad} can be written as
\begin{align}
&\left(1+\bar K\right)\left[\left(\delta\mu_1+\delta\mu_3\right)_{,r}+\left(\frac{1}{r}-\bar\nu_{,r}\right)\left(\delta\mu_1+\delta\mu_3\right)-\frac{2}{r}\delta\mu_2\right]\nonumber\\
&+\,e^{\bar\nu}\left(e^{-\bar\nu}\delta K\right)_{,r}\nonumber\\&+\frac{1}{2}\left(\bar\phi_{,r}\delta\Xi+\bar\Xi_{,r}\delta\phi+\bar\psi_{,r}\delta\zeta+\bar\zeta_{,r}\delta\psi+2\bar\psi\bar\phi_{,r}\delta\phi\right)=0\,.\label{02}
\end{align}
Note that we can drop the derivatives with respect to $x^0$ in the Fourier space.

The $(0,3)$, $(2,3)$, and $(1,1)$ components of the linearized Eq.~\eqref{greqtetrad} read
\begin{align}
\left(1+\bar K\right)&\,\left[\left(\delta\mu_1+\delta\mu_2\right)_{,\theta}+\left(\delta\mu_1-\delta\mu_3\right)\cot\theta\right]=-\delta K_{,\theta}\,,\label{03}\\
\left(1+\bar K\right)\,&\Bigg[\left(\delta\mu_1+\delta\nu\right)_{,r\theta}+\left(\delta\mu_1-\delta\mu_3\right)_{,r}\cot\theta\nonumber\\&+\left(\bar\nu_{,r}-\frac{1}{r}\right)\delta\nu_{,\theta}-\left(\bar\nu_{,r}+\frac{1}{r}\right)\delta\mu_{2,\theta}\Bigg]\nonumber\\
=&-\delta K_{,r\theta}+\frac{1}{r}\delta K_{,\theta}-\frac{1}{2}\big(\bar\phi_{,r}\delta\Xi_{,\theta}+\bar\Xi_{,r}\delta\phi_{,\theta}\nonumber\\&+\bar\psi_{,r}\delta\zeta_{,\theta}+\bar\zeta_{,r}\delta\psi_{,\theta}+2\bar\psi\bar\phi_{,r}\delta\phi_{,\theta}\big)\,,\label{23}
\end{align}
and
\begin{align}
&e^{2\bar\nu}\,\Bigg[\delta\mu_{1,rr}+2\left(\frac{1}{r}+\bar\nu_{,r}\right)\delta\mu_{1,r}\nonumber\\&+\frac{1}{r}\left(\delta\mu_1+\delta\nu+\delta\mu_3-\delta\mu_2\right)_{,r}-\frac{2\delta\mu_2}{r}\left(\frac{1}{r}+2\bar\nu_{,r}\right)\Bigg]\nonumber\\
+&\,\frac{1}{r^2}\left[\delta\mu_{1,\theta\theta}+\cot\theta\left(2\delta\mu_1+\delta\nu+\delta\mu_2-\delta\mu_3\right)_{,\theta}+2\delta\mu_3\right]\nonumber\\&-e^{-2\bar\nu}\delta\mu_{1,tt}
\nonumber\\=&\,\frac{-1}{1+\bar K}\left(\frac{\Box\delta K}{2}+\frac{e^{2\bar\nu}}{r}\delta K_{,r}+\frac{\cot\theta}{r^2}\delta K_{,\theta}\right)\,,\label{11}
\end{align}
respectively. Finally, the $(2,2)$ component of linearized Eq.~\eqref{greqtetrad} can be written as
\begin{align}
e^{2\bar\nu}&\left[\frac{2}{r}\delta\nu_{,r}+\left(\frac{1}{r}+\bar\nu_{,r}\right)\left(\delta\mu_1+\delta\mu_3\right)_{,r}-2\delta\mu_2\left(\frac{1}{r^2}+\frac{2\bar\nu_{,r}}{r}\right)\right]\nonumber\\
&+\frac{1}{r^2}\left[\left(\delta\mu_1+\delta\nu\right)_{,\theta\theta}+\left(2\delta\mu_1+\delta\nu-\delta\mu_3\right)_{,\theta}\cot\theta+2\delta\mu_3\right]\nonumber\\
&-e^{-2\bar\nu}\left(\delta\mu_1+\delta\mu_3\right)_{,tt}\nonumber\\
=&\,\frac{-1}{1+\bar K}\Bigg[\Box\delta K-e^{\bar\nu}\left(\delta K_{,r}e^{\bar\nu}\right)_{,r}-\frac{e^{2\bar\nu}}{2}\big(\bar\phi_{,r}\delta\Xi_{,r}+\delta\phi_{,r}\bar\Xi_{,r}\nonumber\\&+\bar\psi_{,r}\delta\zeta_{,r}+\delta\psi_{,r}\bar\zeta_{,r}+\delta\psi\bar\phi_{,r}^2+2\bar\psi\bar\phi_{,r}\delta\phi_{,r}\big)\Bigg]\,.\label{22}
\end{align}

To proceed further, we consider the following field decompositions \cite{Chandrabook}:
\begin{align}
\delta\nu&=N(r)P_le^{i\omega t}\,,\nonumber\\
\delta\mu_2&=L(r)P_le^{i\omega t}\,,\nonumber\\
\delta\mu_3&=\left[T(r)P_l+V(r)P_{l,\theta\theta}\right]e^{i\omega t}\,,\nonumber\\
\delta\mu_1&=\left[T(r)P_l+V(r)P_{l,\theta}\cot\theta\right]e^{i\omega t}\,,\nonumber\\
\delta\mu_1+\delta\mu_3&=\left[2T-l(l+1)V\right]P_le^{i\omega t}\,,\nonumber\\
\delta K&=(1+\bar K)\delta\tilde{K}(r)P_le^{i\omega t}\,,
\end{align}
where $P_l$ is the Legendre polynomials. Furthermore, according to Eqs.~\eqref{phipsiBGDW2}, \eqref{zetabgdw2dd}, and \eqref{Xibgdw2dd}, we find it convenient to redefine a new scalar field and make the following decomposition
\begin{align}
U&=\frac{\tilde{U}(r)}{r}P_le^{i\omega t}\nonumber\\
&\equiv \frac{C_\phi\delta\Xi+C_\Xi\delta\phi+C_\psi\delta\zeta+\left(C_\phi\bar\phi+C_\zeta\right)\delta\psi}{2\left(1+\bar{K}\right)}\,.\label{decompiU}
\end{align}
Then, Eqs.~\eqref{02}, \eqref{03}, \eqref{23}, \eqref{22} can be written as
\begin{align}
\left[\frac{d}{dr}+\left(\frac{1}{r}-\bar\nu_{,r}\right)\right]&\left[2T-l(l+1)V\right]-\frac{2}{r}L\nonumber\\=&-e^{\bar\nu}\left(e^{-\bar\nu}\delta \tilde{K}\right)_{,r}-\frac{e^{-2\bar\nu}}{r^3}\tilde U\,,\label{37}\\
T-V+L=&-\delta \tilde{K}\,,\label{38}\\
\left(T-V+N\right)_{,r}&-\left(\frac{1}{r}-\bar\nu_{,r}\right)N-\left(\frac{1}{r}+\bar\nu_{,r}\right)L\nonumber\\=&-\delta \tilde{K}_{,r}+\frac{1}{r}\delta \tilde{K}-\frac{e^{-2\bar\nu}}{r^3}\tilde U\,,\label{39}
\end{align}
and
\begin{align}
&e^{2\bar\nu}\Bigg[\frac{2}{r}N_{,r}+\left(\frac{1}{r}+\bar\nu_{,r}\right)\left(2T-l(l+1)V\right)_{,r}\nonumber\\&-\frac{2}{r}\left(\frac{1}{r}+2\bar\nu_{,r}\right)L\Bigg]-\frac{l(l+1)N}{r^2}-\frac{(l+2)(l-1)}{r^2}T\nonumber\\&+e^{-2\bar\nu}\omega^2\left(2T-l(l+1)V\right)\nonumber\\
=&\,-e^{-2\bar\nu}\omega^2\delta \tilde{K}-\left(\frac{2}{r}+\bar\nu_{,r}\right)e^{2\bar\nu}\delta \tilde{K}_{,r}\nonumber\\&+\frac{l(l+1)}{r^2}\delta \tilde{K}+\frac{1}{r^2}\left({\frac{\tilde{U}}{r}}\right)_{,r}\,,\label{40}
\end{align}
respectively.

Defining $X\equiv nV$ where $n\equiv (l+2)(l-1)/2$, one gets
\begin{equation}
2T-l(l+1)V=-2(L+X+\delta\tilde{K})\,,\label{42}
\end{equation}
where we have used Eq.~\eqref{38}. Using Eq.~\eqref{42}, one can rewrite Eqs.~\eqref{37} and \eqref{39} as
\begin{align}
\left(L+X+\frac{\delta\tilde{K}}{2}\right)_{,r}=&-\left(\frac{1}{r}-\bar\nu_{,r}\right)\left(L+X+\frac{\delta\tilde{K}}{2}\right)\nonumber\\-\frac{L}{r}&-\frac{1}{2}\left(\frac{\delta\tilde{K}}{r}-\frac{e^{-2\bar\nu}}{r^3}\tilde U\right)\,,\\
N_{,r}-L_{,r}=\left(\frac{1}{r}-\bar\nu_{,r}\right)N&+\left(\frac{1}{r}+\bar\nu_{,r}\right)L\nonumber\\&+\frac{1}{r}\delta\tilde{K}-\frac{e^{-2\bar\nu}}{r^3}\tilde U\,.
\end{align}
Then, subtracting Eqs.~\eqref{11} from \eqref{22}, and taking only the terms proportional to $P_{l,\theta}\cot\theta$, one gets
\begin{align}
V_{,rr}&+2\left(\frac{1}{r}+\bar\nu_{,r}\right)V_{,r}\nonumber\\&+\frac{e^{-2\bar\nu}}{r^2}\left(N+L+\delta\tilde{K}\right)+\omega^2e^{-4\bar\nu}V=0\,.
\end{align}

Furthermore, one can remove all the $\delta\tilde{K}$ terms by making the following field redefinitions:
\begin{equation}
\tilde{L}\equiv L+\frac{\delta\tilde{K}}{2}\,,\qquad\tilde{N}\equiv N+\frac{\delta\tilde{K}}{2}\,,
\end{equation}
and the above equations can be rewritten as
\begin{align}
&\left(\tilde{L}+X\right)_{,r}=-\left(\frac{1}{r}-\bar\nu_{,r}\right)\left(\tilde{L}+X\right)-\frac{\tilde{L}}{r}+\frac{e^{-2\bar\nu}}{2r^3}\tilde U\,,\label{45b}\\
&\tilde{N}_{,r}-\tilde{L}_{,r}=\left(\frac{1}{r}-\bar\nu_{,r}\right)\tilde{N}+\left(\frac{1}{r}+\bar\nu_{,r}\right)\tilde{L}-\frac{e^{-2\bar\nu}}{r^3}\tilde U\,,\label{46b}\\
&V_{,rr}+2\left(\frac{1}{r}+\bar\nu_{,r}\right)V_{,r}+\frac{e^{-2\bar\nu}}{r^2}\left(\tilde{N}+\tilde{L}\right)+\omega^2e^{-4\bar\nu}V=0\,.\label{47b}
\end{align}
Also, Eq.~\eqref{40} can be written as
\begin{align}
&\frac{2}{r}\tilde{N}_{,r}-2\left(\frac{1}{r}+\bar\nu_{,r}\right)\left(\tilde{L}+X\right)_{,r}-\frac{2}{r}\left(\frac{1}{r}+2\bar\nu_{,r}\right)\tilde{L}\nonumber\\
-&\,\frac{l(l+1)}{r^2}e^{-2\bar\nu}\tilde{N}-\frac{2n}{r^2}\left(V-\tilde{L}\right)e^{-2\bar\nu}-2\omega^2e^{-4\bar\nu}\left(\tilde{L}+X\right)\nonumber\\
=&\,\left(\frac{e^{-2\bar\nu}}{r^2}-\frac{1}{r^2}-\frac{2\bar\nu_{,r}}{r}\right)\delta\tilde{K}+\frac{e^{-2\bar\nu}}{r^2}\left({\frac{\tilde{U}}{r}}\right)_{,r}\nonumber\\
=&\,\frac{e^{-2\bar\nu}}{r^2}\left({\frac{\tilde{U}}{r}}\right)_{,r}\,,\label{48b}
\end{align}
where we have used $e^{2\bar\nu}=1-2m/r$ in the last equality.

Then, using Eqs.~\eqref{45b}, \eqref{46b}, and \eqref{48b}, one can obtain
\begin{align}
\tilde{N}_{,r}=&\,\alpha\tilde{N}+\beta\tilde{L}+\gamma X\nonumber\\&+\frac{e^{-2\bar\nu}}{2r^3}\left[\left(1+r\bar\nu_{,r}\right)\tilde U+r^2\left(\frac{\tilde{U}}{r}\right)_{,r}\right]\,,\\
\tilde{L}_{,r}=&\,\left(\alpha-\frac{1}{r}+\bar\nu_{,r}\right)\tilde{N}+\left(\beta-\frac{1}{r}-\bar\nu_{,r}\right)\tilde{L}+\gamma X\nonumber\\&+\frac{e^{-2\bar\nu}}{2r^3}\left[\left(3+r\bar\nu_{,r}\right)\tilde U+r^2\left(\frac{\tilde{U}}{r}\right)_{,r}\right]\,,\\
X_{,r}=&-\left(\alpha-\frac{1}{r}+\bar\nu_{,r}\right)\tilde{N}-\left(\beta-2\bar\nu_{,r}+\frac{1}{r}\right)\tilde{L}\nonumber\\&-\left(\gamma+\frac{1}{r}-\bar\nu_{,r}\right)X\nonumber\\
&-\frac{e^{-2\bar\nu}}{2r^3}\left[\left(2+r\bar\nu_{,r}\right)\tilde U+r^2\left(\frac{\tilde{U}}{r}\right)_{,r}\right]\nonumber\\
=&\,\frac{\lambda e^{-2\bar\nu}}{r^2}\tilde{N}-\left(\gamma+\frac{1}{r}-\bar\nu_{,r}\right)\left(\tilde{L}+X\right)+\alpha\tilde{L}\nonumber\\
&-\frac{e^{-2\bar\nu}}{2r^3}\left[\left(2+r\bar\nu_{,r}\right)\tilde U+r^2\left(\frac{\tilde{U}}{r}\right)_{,r}\right]\,,
\end{align}
where
\begin{align}
\alpha&=\frac{l(l+1)}{2r}e^{-2\bar\nu}\,,\nonumber\\
\beta&=-\frac{1}{r}+r\left(\bar\nu_{,r}\right)^2+\bar\nu_{,r}-\frac{ne^{-2\bar\nu}}{r}+re^{-4\bar\nu}\omega^2\,,\nonumber\\
\gamma&=-\frac{1}{r}+r\left(\bar\nu_{,r}\right)^2+\frac{e^{-2\bar\nu}}{r}+re^{-4\bar\nu}\omega^2\,,\nonumber\\
\lambda&=-nr-3m\,.
\end{align}

Defining
\begin{equation}
\Psi_{\textrm{polar}}\equiv\frac{r}{n}X+\frac{r^2}{\lambda}\left(\tilde{L}+X\right)\,,
\end{equation}
and differentiating it with respect to the tortoise radius $r_*$ twice, we get the master equation
\begin{align}
\left(\frac{d^2}{dr_*^2}+\omega^2\right)\Psi_{\textrm{polar}}=&\,V_{Z}\Psi_{\textrm{polar}}\nonumber\\+\,e^{2\bar\nu}&\,\left[\frac{3m+2nr}{r^2\left(3m+nr\right)^2}\right]\tilde U\,,\label{polarmaster}
\end{align}
where the Zerilli potential $V_Z$ is \cite{Zerilli:1970se}
\begin{equation}
V_Z=\frac{2e^{2\bar\nu}\left[n^2(n+1)r^3+3mn^2r^2+9m^2nr+9m^3\right]}{r^3\left(nr+3m\right)^2}\,.
\end{equation} 
Therefore, one can see that the polar modes are sourced by the scalar mode $\tilde{U}$. If the scalar mode is not excited, the master equation reduces to the Zerilli equation in GR \cite{Zerilli:1970se}.

From the linearized scalar field equations \eqref{phiboxR}, \eqref{zetaphisquare}, \eqref{Xi2psiphi}, and \eqref{psiRdz}, one finds that the scalar mode $U$ satisfies a massless Klein-Gordon equation{\footnote{It should be mentioned that the constraint Eq.~\eqref{constraintDW2} has to be used in order to prove Eq.~\eqref{4.41boru}.}}:
\begin{equation}
\Box U=0\,,\label{4.41boru}
\end{equation}
which, using the decomposition given by the first line of Eq.~\eqref{decompiU}, can be written as
\begin{equation}
\left(\frac{d^2}{dr_*^2}+\omega^2\right)\tilde U=V_{s}\tilde U\,,
\end{equation}
where the effective potential is
\begin{equation}
V_s=\frac{e^{2\bar\nu}}{r^2}\left[l(l+1)+\frac{2m}{r}\right]\,.
\end{equation}

It should be noted that even if the scalar mode $U$ satisfies the massless Klein-Gordon equation, it is excited only when at least one of the scalar fields ($\bar\phi,\bar\Xi,\bar\psi,\bar\zeta$) is excited already at the background level. More precisely, as one can see from Eq.~\eqref{decompiU}, the scalar mode is excited when there exists at least one non-zero integration constant in the set ($C_\phi,C_\Xi,C_\psi,C_\zeta$). If so, the scalar mode $\tilde{U}$ is dynamical and it would source the polar modes. As a result, the polar modes will be affected by the scalar mode and the isospectrality breaks down. On the other hand, if all the scalar fields are quiescent at the background level ($C_\phi=C_\Xi=C_\psi=C_\zeta=0$), then the scalar mode vanishes and the Schwarzschild black hole in the DW-II nonlocal gravity is completely indistinguishable from its GR counterpart even at the perturbative level.

\section{Remarks for DW-I nonlocal model}\label{sec.dwI}

In this section, we will briefly exhibit that in the DW-I nonlocal gravity, the master equations of the gravitational perturbations share the similar property to those in the DW-II gravity. The resemblance between the two models results from the structural similarity of their equations of motion. In short, the conclusions that the axial gravitational perturbations are the same as in GR, and that the polar gravitational perturbations are sourced by a massless scalar mode, which depends on whether the scalar fields are excited at the background level or not, remain true for the DW-I model. 

The action of the DW-I nonlocal gravity is \cite{Deser:2007jk}
\begin{equation}
\mathcal{S}=\frac{1}{16\pi}\int d^4x\sqrt{-g}R\left[1+f\left(\Box^{-1}R\right)\right]+\mathcal{S}_m\,,\label{action1}
\end{equation}
Following the procedure in Ref.~\cite{Nojiri:2007uq}, one can localize the gravitational action \eqref{action1} by introducing an auxiliary scalar field $\phi$, as in Eq.~\eqref{defphi}, and a Lagrange multiplier $\xi$, such that the action can be written as
\begin{equation}
\mathcal{S}=\frac{1}{16\pi}\int d^4x\sqrt{-g}\left[R\left(1+f\right)-\partial_\mu\xi\partial^\mu\phi-\xi R\right]+\mathcal{S}_m\,.\label{action2}
\end{equation}

The equations of motion are derived by varying the action \eqref{action2} with respect to $\xi$, $\phi$, and the metric $g_{\mu\nu}$. Varying the action with respect to the Lagrange multiplier $\xi$ essentially gives Eq.~\eqref{phiboxR}. Then, after varying the action with respect to the auxiliary scalar field $\phi$, one can obtain
\begin{equation}
\Box\xi=-R\frac{df}{d\phi}\,.\label{boxxi}
\end{equation}
Finally, the variation of the action \eqref{action2} with respect to the metric $g_{\mu\nu}$ leads to the modified Einstein field equation
\begin{equation}
G_{\mu\nu}+\Delta G_{\mu\nu}=8\pi T_{\mu\nu}\,,\label{modifiedeinsteineq}
\end{equation}
where
\begin{align}
\Delta G_{\mu\nu}=&\,\left(G_{\mu\nu}+g_{\mu\nu}\Box-\nabla_\mu\nabla_\nu\right)\left(f-\xi\right)\nonumber\\&+\frac{1}{2}g_{\mu\nu}\partial_\rho\xi\partial^\rho\phi-\partial_{(\mu}\xi\partial_{\nu)}\phi\,.
\end{align}
 
For the non-perturbed Schwarzschild metric with the Ricci scalar being zero, one gets
\begin{equation}
\bar\phi_{,r}=C_\phi\frac{e^{-2\bar\nu}}{r^2}\,,\qquad \bar\xi_{,r}=C_\xi\frac{e^{-2\bar\nu}}{r^2}\,,
\end{equation}
where $C_\xi$ is an integration constant. Furthermore, at the background level, the gravitational equation \eqref{modifiedeinsteineq} implies that
\begin{equation}
\bar F\equiv\bar f-\bar\xi=\textrm{constant}\,, \quad C_\phi C_\xi=0\,.
\end{equation}
Therefore, in the DW-I model, at least one of the scalar fields ($\bar\phi$ and $\bar\xi$) has to be a constant. 

For the perturbations of the Schwarzschild black hole, it is easy to find that the master equation of the axial perturbations is again the Regge-Wheeler equation
\begin{equation}
\left(\frac{d^2}{dr_*^2}+\omega^2\right)\Psi_{\textrm{aixal}}=V_{RW}\Psi_{\textrm{aixal}}\,,\label{axialmaster2}
\end{equation}
where the effective potential $V_{RW}$ is
\begin{equation}
{V_{RW}=\frac{e^{2\bar\nu}}{r^2}\left[l(l+1)-\frac{6m}{r}\right]}\,,\quad e^{2\bar\nu}=1-\frac{2m}{r}\,.
\end{equation}

As for the polar perturbations, we find that upon defining the following scalar mode 
\begin{equation}
W\equiv\frac{\tilde{W}(r)}{r}P_le^{i\omega t}\equiv\frac{1}{2\left(1+\bar{F}\right)}\left(C_\phi\delta\xi+C_\xi\delta\phi\right)\,,
\label{decompiW}
\end{equation}
where we have defined $F\equiv f-\xi$ and decompose $F$ as $F=\bar{F}+\delta F$, the master equation of the polar perturbations in the DW-I gravity can be derived under the recipe of the DW-II gravity, with the replacement $(\bar K,\delta K,\tilde U)\rightarrow(\bar F,\delta F,\tilde W)$. Therefore, the master equation of the polar perturbations takes the following form:
\begin{align}
\left(\frac{d^2}{dr_*^2}+\omega^2\right)\Psi_{\textrm{polar}}=&\,V_{Z}\Psi_{\textrm{polar}}\nonumber\\+\,e^{2\bar\nu}&\,\left[\frac{3m+2nr}{r^2\left(3m+nr\right)^2}\right]\tilde W\,,\label{polarmasterDW1op}
\end{align}
where the Zerilli potential $V_Z$ is given by
\begin{equation}
V_Z=\frac{2e^{2\bar\nu}\left[n^2(n+1)r^3+3mn^2r^2+9m^2nr+9m^3\right]}{r^3\left(nr+3m\right)^2}\,.
\end{equation} 
Similar to what we have done in the DW-II nonlocal gravity, by using the linearized Eqs.~\eqref{phiboxR}, \eqref{boxxi}, and the constraint $C_\phi C_\xi=0$, it can be proven that $\Box W=0$. Therefore, the polar modes are sourced by a massless scalar mode $W$, which is defined by Eq.~\eqref{decompiW}. One can see that whether the scalar mode is excited or not depends on whether the scalar fields $(\phi,\xi)$ are excited or not at the background level. More specifically, if none of them is excited at the background level ($C_\phi=C_\xi=0$), the polar gravitational perturbations are completely the same as those in GR. On the other hand, if one of the integration constants ($C_\phi,C_\xi$) is not zero, the polar modes would be sourced by the massless scalar mode $W$, and the isospectrality would break.

\section{Conclusions}\label{sec.conclus}

The DW-I nonlocal gravity was originally proposed in order to explain the late-time acceleration of the universe without resorting to any mysterious dark energy. The recent discovery \cite{Belgacem:2018wtb} that the DW-I is conflict with the solar system tests has prompted the authors of Ref.~\cite{Deser:2019lmm} to improve the model and propose the DW-II nonlocal gravity. These two theories share an interesting property, which is that the auxiliary scalar fields in the localized framework are not completely dynamical, but subject to the initial data. It is thus interesting to see whether these nonlocal corrections would alter the QNMs or not, as compared with those in GR.

In this paper, we investigated the gravitational perturbations of the Schwarzschild black hole in the DW-II nonlocal gravity. The gravitational perturbations of the Schwarzschild black hole in GR are governed by the Regge-Wheeler equation (axial modes) and the Zerilli equation (polar modes), and the two corresponding modes share the same QNM spectrum. In fact, the isospectrality between the axial and the polar modes have been shown to be very fragile \cite{Cardoso:2019mqo}, in the sense that it easily breaks when the underlying theory is slightly changed. For instance, the isospectrality is no longer satisfied in the presence of a dynamical scalar field, such as in the $f(R)$ gravity \cite{Bhattacharyya:2017tyc,Datta:2019npq}.

Essentially, in this work we showed that the axial gravitational perturbations in the DW-II gravity are completely identical with those in GR. On the other hand, the polar modes are sourced by a massless scalar mode, if there is at least one auxiliary scalar field being excited at the background level. However, if all the auxiliary scalar fields stay quiescent at the background level, the scalar mode disappears and the polar modes reduce to those of the Schwarzschild black hole in GR. Whether the auxiliary fields are excited or not depends on the values of their corresponding integration constants ($C_\phi,C_\psi,C_\Xi,C_\zeta$). We also performed a similar analysis in the DW-I gravity and, thanks to their structural resemblance, a similar conclusion was drawn.

Remarkably, although the excitation of the auxiliary scalar fields does break the isospectrality, meaning that the observations of isospectrality breaking can constrain the integration constants of the auxiliary scalar fields, however, the question about whether one can really falsify the DW-II gravity via isospectrality tests should be answered with great care. The reason is that what one really constrains, through isospectrality tests, are just some integration constants which do not directly characterize the theory. This is different from the cases in some other models (e.g. $f(R)$ gravity or dynamical Chern-Simons gravity) in which the isospectrality tests directly constrain the coupling constants of the theories. This thus makes the DW-II gravity more intrinsically tenacious against isospectrality tests. Therefore, at this point, since DW-I model has been ruled out, we want to point out that the DW-II nonlocal gravity seems not only successful in describing the accelerating expansion of our universe in the cosmological scale, but also theoretically consistent down to small scales as such of black holes. From an astrophysical point of view, to confirm the validity of the DW model on a more concrete basis, one should then consider rotating black holes since astrophysical black holes are usually spinning. It would be interesting to see in such cases, whether the DW model still gives predictions identical with those made in GR, or some distinctive features would emerge. We leave these issues for our future works.

\section*{Acknowledgments}
CYC is supported by the Institute of Physics of Academia Sinica.


\section*{References}

\begin{thebibliography}{99}

\bibitem{Abbott:2016blz}
B.~P.~Abbott \textit{et al.} [LIGO Scientific and Virgo],
Phys. Rev. Lett. \textbf{116} (2016) no.6, 061102.

\bibitem{LIGOScientific:2018mvr}
B.~P.~Abbott \textit{et al.} [LIGO Scientific and Virgo],
Phys. Rev. X \textbf{9} (2019) no.3, 031040.

\bibitem{Abbott:2020niy}
R.~Abbott \textit{et al.} [LIGO Scientific and Virgo],
[arXiv:2010.14527 [gr-qc]].

\bibitem{Kokkotas:1999bd}
K.~D.~Kokkotas and B.~G.~Schmidt,
Living Rev. Rel. \textbf{2} (1999), 2.

\bibitem{Berti:2009kk}
E.~Berti, V.~Cardoso and A.~O.~Starinets,
Class. Quant. Grav. \textbf{26}, 163001 (2009).

\bibitem{Konoplya:2011qq}
R.~A.~Konoplya and A.~Zhidenko,
Rev. Mod. Phys. \textbf{83} (2011), 793-836.

\bibitem{Kobayashi:2012kh}
T.~Kobayashi, H.~Motohashi and T.~Suyama,
Phys. Rev. D \textbf{85}, 084025 (2012)
[erratum: Phys. Rev. D \textbf{96}, no.10, 109903 (2017)].

\bibitem{Kobayashi:2014wsa}
T.~Kobayashi, H.~Motohashi and T.~Suyama,
Phys. Rev. D \textbf{89}, no.8, 084042 (2014).

\bibitem{Blazquez-Salcedo:2016enn}
J.~L.~Bl\'azquez-Salcedo, C.~F.~B.~Macedo, V.~Cardoso, V.~Ferrari, L.~Gualtieri, F.~S.~Khoo, J.~Kunz and P.~Pani,
Phys. Rev. D \textbf{94}, no.10, 104024 (2016).

\bibitem{Bhattacharyya:2017tyc}
S.~Bhattacharyya and S.~Shankaranarayanan,
Phys. Rev. D \textbf{96} (2017) no.6, 064044.

\bibitem{Glampedakis:2017dvb}
K.~Glampedakis, G.~Pappas, H.~O.~Silva and E.~Berti,
Phys. Rev. D \textbf{96}, no.6, 064054 (2017).

\bibitem{Chen:2018mkf}
C.~Y.~Chen and P.~Chen,
Phys. Rev. D \textbf{98}, no.4, 044042 (2018).

\bibitem{Chen:2018vuw}
C.~Y.~Chen, M.~Bouhmadi-L\'opez and P.~Chen,
Eur. Phys. J. C \textbf{79}, no.1, 63 (2019).

\bibitem{Chen:2019iuo}
C.~Y.~Chen and P.~Chen,
Phys. Rev. D \textbf{99}, no.10, 104003 (2019).

\bibitem{Moulin:2019ekf}
F.~Moulin, A.~Barrau and K.~Martineau,
Universe \textbf{5}, no.9, 202 (2019).

\bibitem{Chen:2020evr}
C.~Y.~Chen, Y.~H.~Kung and P.~Chen,
Phys. Rev. D \textbf{102}, 124033 (2020).

\bibitem{Cardoso:2008bp}
V.~Cardoso, A.~S.~Miranda, E.~Berti, H.~Witek and V.~T.~Zanchin,
Phys. Rev. D \textbf{79}, 064016 (2009).

\bibitem{Dolan:2010wr}
S.~R.~Dolan,
Phys. Rev. D \textbf{82}, 104003 (2010).

\bibitem{Yang:2012he}
H.~Yang, D.~A.~Nichols, F.~Zhang, A.~Zimmerman, Z.~Zhang and Y.~Chen,
Phys. Rev. D \textbf{86}, 104006 (2012).

\bibitem{Cuadros-Melgar:2020kqn}
B.~Cuadros-Melgar, R.~D.~B.~Fontana and J.~de Oliveira,
Phys. Lett. B \textbf{811}, 135966 (2020).



\bibitem{Konoplya:2017wot}
R.~A.~Konoplya and Z.~Stuchl\'\i{}k,
Phys. Lett. B \textbf{771}, 597-602 (2017).

\bibitem{Churilova:2019jqx}
M.~S.~Churilova,
Eur. Phys. J. C \textbf{79}, no.7, 629 (2019).

\bibitem{Glampedakis:2019dqh}
K.~Glampedakis and H.~O.~Silva,
Phys. Rev. D \textbf{100}, no.4, 044040 (2019).

\bibitem{Chen:2019dip}
C.~Y.~Chen and P.~Chen,
Phys. Rev. D \textbf{101}, no.6, 064021 (2020).

  \bibitem{Chandrabook}
S.~Chandrasekhar(ed.): The Mathematical Theory of Black Holes. Oxford University Press,
Oxford (1992).

\bibitem{Pani:2013ija}
P.~Pani, E.~Berti and L.~Gualtieri,
Phys. Rev. Lett. \textbf{110}, no.24, 241103 (2013).

\bibitem{Pani:2013wsa}
P.~Pani, E.~Berti and L.~Gualtieri,
Phys. Rev. D \textbf{88}, 064048 (2013).






\bibitem{Glampedakis:2017rar}
K.~Glampedakis, A.~D.~Johnson and D.~Kennefick,
Phys. Rev. D \textbf{96}, no.2, 024036 (2017).




\bibitem{Darboux:1882}
 G. Darboux, 
 C. R. Academy Sci. (Paris) \textbf{94}, 1456 (1882).




\bibitem{Yurov:2018ynn}
A.~V.~Yurov and V.~A.~Yurov,
Phys. Lett. A \textbf{383}, no.22, 2571-2578 (2019).



\bibitem{Teukolsky:1973ha}
S.~A.~Teukolsky,
Astrophys. J. \textbf{185}, 635-647 (1973).

\bibitem{Sasaki:1981sx}
M.~Sasaki and T.~Nakamura,
Prog. Theor. Phys. \textbf{67}, 1788 (1982).


\bibitem{Moulin:2019bfh}
F.~Moulin and A.~Barrau,
Gen. Rel. Grav. \textbf{52}, no.8, 82 (2020).



\bibitem{Ferrari:2000ep}
V.~Ferrari, M.~Pauri and F.~Piazza,
Phys. Rev. D \textbf{63}, 064009 (2001).


\bibitem{Cardoso:2001bb}
V.~Cardoso and J.~P.~S.~Lemos,
Phys. Rev. D \textbf{64}, 084017 (2001).

\bibitem{Berti:2003ud}
E.~Berti and K.~D.~Kokkotas,
Phys. Rev. D \textbf{67}, 064020 (2003).

\bibitem{Blazquez-Salcedo:2019nwd}
J.~L.~Bl\'azquez-Salcedo, S.~Kahlen and J.~Kunz,
Eur. Phys. J. C \textbf{79}, no.12, 1021 (2019).






\bibitem{Bhattacharyya:2018qbe}
S.~Bhattacharyya and S.~Shankaranarayanan,
Eur. Phys. J. C \textbf{78} (2018) no.9, 737.

\bibitem{Datta:2019npq}
S.~Datta and S.~Bose,
Eur. Phys. J. C \textbf{80} (2020) no.1, 14.


\bibitem{Bhattacharyya:2018hsj}
S.~Bhattacharyya and S.~Shankaranarayanan,
Phys. Rev. D \textbf{100}, no.2, 024022 (2019).


\bibitem{Cardoso:2019mqo}
V.~Cardoso, M.~Kimura, A.~Maselli, E.~Berti, C.~F.~B.~Macedo and R.~McManus,
Phys. Rev. D \textbf{99}, no.10, 104077 (2019).

\bibitem{Shankaranarayanan:2019yjx}
S.~Shankaranarayanan,
Int. J. Mod. Phys. D \textbf{28}, no.14, 1944020 (2019).

\bibitem{Deser:2007jk}
  S.~Deser and R.~P.~Woodard,
  Phys.\ Rev.\ Lett.\  {\bf 99} (2007) 111301.


\bibitem{Deser:2019lmm}
S.~Deser and R.~P.~Woodard,
JCAP \textbf{06}, 034 (2019).
  



\bibitem{Woodard:2014iga}
R.~P.~Woodard,
Found. Phys. \textbf{44} (2014), 213.


\bibitem{Woodard:2018gfj}
R.~P.~Woodard,
Universe \textbf{4} (2018), 
88.


\bibitem{Barvinsky:2011hd}
  A.~O.~Barvinsky,
  Phys.\ Lett.\ B {\bf 710} (2012) 12.
  
  
\bibitem{Maggiore:2013mea}
  M.~Maggiore,
  Phys.\ Rev.\ D {\bf 89} (2014) 043008.

\bibitem{Maggiore:2014sia}
  M.~Maggiore and M.~Mancarella,
  Phys.\ Rev.\ D {\bf 90} (2014) 023005.

\bibitem{Vardanyan:2017kal}
  V.~Vardanyan, Y.~Akrami, L.~Amendola and A.~Silvestri,
  JCAP {\bf 1803} (2018) 048.

\bibitem{Amendola:2017qge}
  L.~Amendola, N.~Burzilla and H.~Nersisyan,
  Phys.\ Rev.\ D {\bf 96} (2017) 084031.

\bibitem{Tian:2018bmn}
  S.~Tian,
  Phys.\ Rev.\ D {\bf 98} (2018) 084040.
  
\bibitem{Deffayet:2009ca}
C.~Deffayet and R.~P.~Woodard,
JCAP \textbf{08} (2009), 023.

\bibitem{Park:2012cp}
  S.~Park and S.~Dodelson,
  Phys.\ Rev.\ D {\bf 87} (2013) 024003.

\bibitem{Dodelson:2013sma}
  S.~Dodelson and S.~Park,
  Phys.\ Rev.\ D {\bf 90} (2014) 043535,
   Erratum: [Phys.\ Rev.\ D {\bf 98} (2018) 029904].
  
\bibitem{Park:2016jym}
  S.~Park and A.~Shafieloo,
  Phys.\ Rev.\ D {\bf 95} (2017) 064061.
  
\bibitem{Nersisyan:2017mgj}
H.~Nersisyan, A.~F.~Cid and L.~Amendola,
JCAP \textbf{04} (2017), 046.
  
    
\bibitem{Park:2017zls}
  S.~Park,
  Phys.\ Rev.\ D {\bf 97} (2018) 044006.
  
\bibitem{Amendola:2019fhc}
L.~Amendola, Y.~Dirian, H.~Nersisyan and S.~Park,
JCAP \textbf{03} (2019), 045.

\bibitem{Nojiri:2007uq}
S.~Nojiri and S.~D.~Odintsov,
Phys. Lett. B \textbf{659}, 821-826 (2008).

\bibitem{Zhang:2016ykx}
  Y.~l.~Zhang, K.~Koyama, M.~Sasaki and G.~B.~Zhao,
  JHEP {\bf 1603} (2016) 039.

\bibitem{Nojiri:2010pw}
  S.~Nojiri, S.~D.~Odintsov, M.~Sasaki and Y.~l.~Zhang,
  Phys.\ Lett.\ B {\bf 696} (2011) 278.

\bibitem{Zhang:2011uv}
  Y.~l.~Zhang and M.~Sasaki,
  Int.\ J.\ Mod.\ Phys.\ D {\bf 21} (2012) 1250006.


\bibitem{Park:2019btx}
  S.~Park and R.~P.~Woodard,
  Phys.\ Rev.\ D {\bf 99} (2019) 024014.
 
 
\bibitem{Belgacem:2018wtb}
  E.~Belgacem, A.~Finke, A.~Frassino and M.~Maggiore,
  JCAP {\bf 1902} (2019) 035.
 




\bibitem{Chu:2018mld}
  Y.~Z.~Chu and S.~Park,
  Phys.\ Rev.\ D {\bf 99} (2019) 044052.


\bibitem{Regge:1957td}
  T.~Regge and J.~A.~Wheeler,
  Phys.\ Rev.\  {\bf 108} (1957) 1063.

\bibitem{Zerilli:1970se}
F.~J.~Zerilli,
Phys. Rev. Lett. \textbf{24}, 737-738 (1970).
















\end{thebibliography}
\end{document}